\documentclass[american,aps,prb,reprint,groupedaddress,superscriptaddress]{revtex4-1}
\usepackage[T1]{fontenc}
\usepackage[utf8]{inputenc}
\usepackage{color}
\usepackage{amsmath}
\usepackage{graphicx}
\usepackage{esint}

\makeatletter
\usepackage{mciteplus}
\usepackage{hyphenat}

\usepackage{babel}

\makeatother

\usepackage{babel}
\begin{document}

\title{First-principles calculation of the thermoelectric figure of merit
for {[}2,2{]}paracyclophane-based single-molecule junctions}

\author{Marius Bürkle}

\email{marius.buerkle@aist.go.jp}

\affiliation{Nanosystem Research Institute (NRI) ‘RICS’, National Institute of
Advanced Industrial Science and Technology (AIST), Umezono 1-1-1,
Tsukuba Central 2, Tsukuba, Ibaraki 305-8568, Japan}

\author{Thomas J. Hellmuth}

\affiliation{Institut für Theoretische Festkörperphysik, Karlsruhe Institute of
Technology, 76131 Karlsruhe, Germany}

\affiliation{Department of Physics, University of Konstanz, 78457 Konstanz, Germany}

\author{Fabian Pauly}

\affiliation{Department of Physics, University of Konstanz, 78457 Konstanz, Germany}

\author{Yoshihiro Asai}

\affiliation{Nanosystem Research Institute (NRI) ‘RICS’, National Institute of
Advanced Industrial Science and Technology (AIST), Umezono 1-1-1,
Tsukuba Central 2, Tsukuba, Ibaraki 305-8568, Japan}
\begin{abstract}
Here we present a theoretical study of the thermoelectric transport
through {[}2,2{]}para\-cyclo\-phane-based single-molecule junctions.
Combining electronic and vibrational structures, obtained from density
functional theory (DFT), with nonequilibrium Green's function techniques,
allows us to treat both electronic and phononic transport properties
at a first-principles level. For the electronic part, we include an
approximate self-energy correction, based on the DFT+$\Sigma$ approach.
This enables us to make a reliable prediction of all linear response
transport coefficients entering the thermoelectric figure of merit
$ZT$. Paracyclophane derivatives offer a great flexibility in tuning
their chemical properties by attaching different functional groups.
We show that, for the specific molecule, the functional groups mainly
influence the thermopower, allowing to tune its sign and absolute
value. We predict that the functionalization of the bare paracyclophane
leads to a largely enhanced electronic contribution $Z_{\mathrm{el}}T$
to the figure of merit. Nevertheless, the high phononic contribution
to the thermal conductance strongly suppresses $ZT$. Our work demonstrates
the importance to include the phonon thermal conductance for any realistic
estimate of the $ZT$ for off-resonant molecular transport junctions.
In addition, it shows the possibility of a chemical tuning of the
thermoelectric properties for a series of available molecules, leading
to equally performing hole- and electron-conducting junctions based
on the same molecular framework. 
\end{abstract}
\maketitle

\section{Introduction}

Nanostructured materials provide a promising route towards high-performance
thermoelectric energy conversion.\cite{Hicks1993,Bell2008,Kanatzidis2010}
Nanostructuring does not only allow to improve the thermoelectric
performance, but also enables novel applications like on-chip solutions
for energy harvesting and refrigeration.\cite{Bell2008} Recently
it became feasible to probe the thermoelectric properties and energy
conversion in molecular junctions.\cite{Reddy2007,Baheti2008,Lee2013}
Within the different approaches to novel thermoelectric materials,
molecular junctions are the ultimate level achievable regarding device
miniaturization and provide a great flexibility to tailor their thermoelectric
properties, e.g.~by chemical synthesis. Presently the applicability
of organic thermoelectric materials is limited by the absence of high
performance electron-type materials.\cite{Zhang2014} The possibility
to realize equally well performing hole- and electron-type ($p$-
and $n$-type) molecular junctions and to enhance their efficiency\cite{Yee2011,Evangeli2013,Widawsky2013}
makes them very appealing for further scientific investigation. Irrespective
of their practical use, they represent ideal systems to study thermoelectric
transport at the atomic scale and to improve the understanding of
the fundamental processes of nanoscale energy conversion.

On the theory side the combination of DFT-based electronic structure
calculations with nonequilibrium Green's function techniques allows
an atomistic description of the electric and electronic thermal transport
properties. With this approach important trends like the influence
of the chemical and structural properties on the charge transport
can be successfully captured.\cite{Venkataraman2006,Reddy2007,Pauly:PRB2008,Mishchenko:NanoLett2009}
Moreover the recently introduced DFT+$\Sigma$ method, which improves
the description of the level alignment at the metal-molecule interface,
yields a quantitative agreement with experimental results at low computational
costs.\cite{Quek2007}

Recent computational studies on molecular junctions\cite{Finch2009,Bergfield2010,Stadler2011}
suggest that antiresonances in the electronic transmission close to
the Fermi energy can largely enhance the thermopower. The decrease
of the electric conductance will be compensated by a simultaneous
decrease of the electronic contribution to the thermal conductance.
Consequently, this is reported to lead to an increase of the electronic
contribution to the thermoelectric figure of merit $Z_{\mathrm{el}}T$.
In such situations the overall thermoelectric figure of merit $ZT$
is eventually limited by the phononic contribution to the thermal
conductance (see Eqs.~(\ref{eq:ZT_def}) and (\ref{eq:ZTel_def})
for definitions of $ZT_{\mathrm{el}}$ and $ZT$). In Refs.~\citenum{Finch2009,Bergfield2010,Stadler2011},
however, this phonon thermal conductance is either neglected or just
taken into account in an approximate manner. While it is speculated
in Ref.~\citenum{Bergfield2010} that the phonon thermal conductance
might just play a minor role, it is suggested in Ref.~\citenum{Stadler2011}
that it can have a significant contribution to the overall thermal
conductance. Based on first-principles calculations of the relevant
transport coefficients, we demonstrate in this work that it is indispensable
for an accurate description of the energy conversion capabilities
of molecular junctions to include also the phononic contribution.

As an accurate total energy method, DFT provides the ideal tool to
study phonon transport from first-principles and has already been
applied successfully to various bulk and nanostructured systems.\cite{Ward2009,Calzolari2012,Luckyanova2012}
However, for molecular junctions first-principles calculations of
the phonon thermal conductance have up to now only been performed
for a few idealized systems.\cite{Nozaki2010,Nikolic2012,Nakamura2013,Asai2013}
Systematic studies for realistic single-molecule junctions have yet
to be provided.

We present here a fully first-principles based study of the electron
and phonon transport properties for a series of paracyclophane derivatives
connected to gold electrodes. We focus on the question, how charge
and phonon transport can be tuned chemically by introducing different
side-groups, which are either strongly electron-withdrawing or electron-donating.
The chemical structure of the considered {[}2,2{]}para\-cyclo\-phane
derivatives is given in Fig.~\ref{fig:mols_geos}a, namely {[}2,2{]}paracyclophane,
diamino\hyp{}{[}2,2{]}paracyclophane, dihydroxy\hyp{}{[}2,2{]}paracyclophane,
dinitro\hyp{[}2,2{]}paracyclophane and ditrifluoroacetyl\hyp{}{[}2,2{]}paracyclophane,
where we, for simplicity, just analyze the pseudo-para isomers. For
{[}2,2{]}para\-cyclo\-phane the single-molecule electrical conductance
was measured recently.\cite{Schneebeli2011}

\section{Formalism and methods}

\subsection{Electronic, geometric and vibrational structure }

\begin{figure}[!tb]
\begin{centering}
\includegraphics[width=1\linewidth]{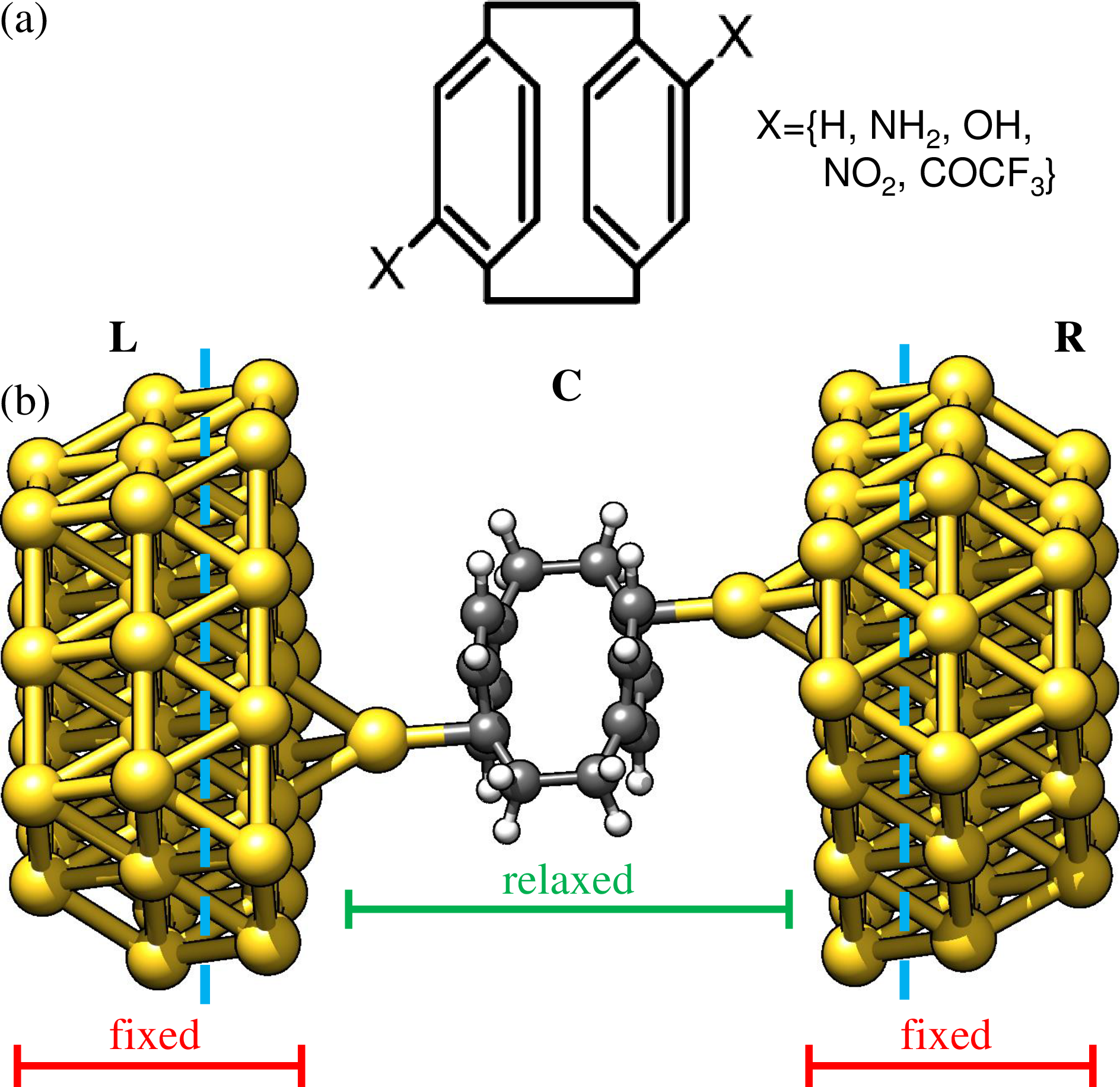} 
\par\end{centering}

\protect\protect\protect\protect\caption{(a) Investigated paracyclophane derivatives with $\mathsf{X=\{H,\,NH_{2},\,OH,\,NO_{2},\,COCF_{3}\}}$
in pseudo-para position. (b) Example of the used contact geometry,
here with $\mathsf{X=H}$. \label{fig:mols_geos}}
\end{figure}

The electronic structure and the contact geometries are obtained at
the DFT level, and the vibrational properties in the framework of
density functional perturbation theory (DFPT). Both procedures are
implemented in the quantum chemistry software package TURBOMOLE 6.4.\cite{TURBOMOLE6.4,Deglmann2002,Deglmann2004}
We employ the PBE functional\cite{Dirac1929,Slater1951,Perdew1992,Perdew1996}
with empirical dispersion corrections to the total ground-state energy,
accounting for long-range van der Waals interactions.\cite{Grimme2006}
As basis set def2-SV(P)\cite{Weigend2005} and the respective Coulomb
fitting basis\cite{Weigend2006} are used. To ensure an accurate description
of the vibrational properties, we use very tight convergence criteria.
Total energies are converged to a precision of better than $10^{-9}$
a.u., while geometry optimizations are performed until the change
of the maximum norm of the Cartesian gradient is below $10^{-5}$
a.u.. 

In the construction of the contact geometries, as shown in Fig.~\ref{fig:mols_geos}b,
we assume the \textsf{Au} atoms of the electrodes in (111) orientation
to be fixed at their ideal fcc lattice positions (lattice constant
$a_{0}=4.08$~Å). In contrast, the two \textsf{Au} apex atoms and
the molecule are fully relaxed. While the precise metal-molecule-metal
geometry will influence the thermoelectric properties, we focus here
on the effect of the substituents. Due to the time-consuming DFPT
calculations, we therefore restrict ourselves to one representative
contact configuration. This contact geometry or extended central cluster
(ECC)\cite{Pauly2008} and its division into left (L), central (C)
and right (R) regions is shown in Fig.~\ref{fig:mols_geos}b for
the bare, unsubstituted {[}2,2{]}paracyclophane.

\subsection{Electronic transport}

In the Landauer-Büttiker picture the charge current $I$ and the electronic
thermal current $Q_{\mathrm{el}}$ are given by\cite{Datt1997,Sivan1986,Zotti2014}
\begin{alignat}{1}
I & =\dfrac{2e}{h}\int\mbox{d}E\tau_{\textrm{el}}(E)\left[f(E,\mu_{\mathrm{L}},T_{\mathrm{L}})\right.\nonumber \\
 & \left.-f(E,\mu_{\mathrm{R}},T_{\mathrm{R}})\right],\\
Q_{\mathrm{el}} & =\dfrac{2}{h}\int\mbox{d}E(E-\mu)\tau_{\mathrm{el}}(E)\times\nonumber \\
 & \left[f(E,\mu_{\mathrm{L}},T_{\mathrm{L}})-f(E,\mu_{\mathrm{R}},T_{\mathrm{R}})\right].\label{eq:Qel-1}
\end{alignat}
Here, $\tau_{\textrm{el}}(E)$ is the energy-dependent electron transmission
(e.g.~from L through C into R, see Fig.~\ref{fig:mols_geos}b) and
$f(E,\mu,T)=\left\{ \exp[(E-\mu)/k_{\mathrm{B}}T]+1\right\} ^{-1}$
is the Fermi function, describing the occupation of the left (right)
electron reservoir at the chemical potential $\mu_{\textrm{L}}$ ($\mu_{\mathrm{R}}$)
and the temperature $T_{\textrm{L}}$ ($T_{\textrm{R}}$). The energy
of the electrons is measured relative to the average chemical potential
$\mu=(\mu_{\mathrm{L}}+\mu_{\mathrm{R}})/2$. This is valid in the
linear response regime, assuming $\Delta\mu=\mu_{\textrm{L}}-\mu_{\textrm{R}}$
and $\Delta T=T_{\textrm{L}}-T_{\textrm{R}}$ to be infinitesimally
small.

In linear response the thermoelectric transport coefficients, that
is the electrical conductance $G$, thermopower $S$ and electron
thermal conductance $\kappa_{\textnormal{el}}$, can be expressed
as follows\cite{Sivan1986,Esfarjani2006,Mueller2008}

\begin{align}
G & =\frac{2e^{2}}{h}K_{0},\label{eq:G}\\
S & =-\dfrac{K_{1}}{eTK_{0}},
\end{align}
\begin{align}
\kappa_{\mathrm{el}} & =\dfrac{2}{hT}\left(K_{2}-\dfrac{K_{1}^{2}}{K_{0}}\right).\label{eq:kel}
\end{align}
In the expressions $e=\left|e\right|$ is the absolute value of the
electron charge, $h$ is the Planck constant, $k_{\mathrm{B}}$ is
the Boltzmann constant, and $T=(T_{\mathrm{L}}+T_{\mathrm{R}})/2$
is the average junction temperature. The coefficients in Eqs.~(\ref{eq:G})-(\ref{eq:kel})
are defined as 
\begin{equation}
K_{n}=\int\mbox{d}E\tau_{\mathrm{el}}(E)\left(-\tfrac{\partial f(E)}{\partial E}\right)(E-\mu)^{n},\label{eq:Kn}
\end{equation}
where the chemical potential $\mu\approx E_{\textnormal{F}}$ is approximately
given by the Fermi energy $E_{\textnormal{F}}$ of the \textsf{Au}
electrodes.

The electronic transmission function is expressed within the Green's
function formalism.\cite{Datt1997,Pauly2008} The quasi-particle energies,
entering the electronic Green's functions, are approximated by the
Kohn-Sham eigenvalues, including a ``self-energy correction''. This
self-energy correction of the DFT+$\Sigma$ method \cite{Quek2007}
consists of two parts, one that accounts for the underestimation of
the electronic gap of the gas-phase molecule in local and semilocal
approximations to DFT, the other one for long-range correlations in
a junction that stabilize the charge on the molecule through image
charges in the electrodes. Details regarding the self-energy correction
can be found in Ref.~\citenum{Zotti2014}. The electronic contact
self-energies, on the other hand, are obtained as described in Ref.~\citenum{Pauly2008}.
To construct the electrode surface Green's function, we use $64\times64$
$k$-points in the transverse direction, which was found to be sufficient
to obtain converged electronic transport coefficients.

\subsection{Phononic transport\label{sub:Phonon-transport}}

Since phonons are chargeless, they can only carry a heat current.
Similar to the electronic expression, this phononic heat current can
be calculated in the Landauer-Büttiker picture\cite{Rego1998,Mingo2003,Yamamoto2006}
as
\begin{equation}
Q_{\mathrm{ph}}=\dfrac{1}{h}\int_{0}^{\infty}\mbox{d}EE\tau_{\textrm{ph}}(E)\left[n(E,T_{\mathrm{L}})-n(E,T_{\mathrm{R}})\right].\label{eq:Qph}
\end{equation}
The corresponding thermal conductance due to the phonons is given
in linear response by 
\begin{equation}
\kappa_{\mathrm{ph}}=\dfrac{1}{h}\int_{0}^{\infty}\mbox{d}EE\tau_{\textrm{ph}}(E)\dfrac{\partial n(E,T)}{\partial T},\label{eq:kph}
\end{equation}
where $\tau_{\textrm{ph}}(E)$ is the phonon transmission and $n(E,T)=\{\exp(E/k_{\mathrm{B}}T)-1\}^{-1}$
is the Bose function, characterizing the phonon reservoirs in the
left and right electrodes.

Similar to the electronic transmission function, those of the phonons
can also be computed within the Green's function formalism.\cite{Datt1997,Mingo2003,Wang2006,Yamamoto2006,Asai2008}
The system of interest here, namely a molecular junction or more generally
a nanoscale conductor, consists of an arbitrary, but finite-size scattering
region (C), connected to two semi-infinite electrodes (L and R). For
small displacements $\{Q_{\xi}\}$ around the atomic equilibrium positions
$\{R_{\xi}^{(0)}\}$ the phononic Hamiltonian in the harmonic approximation
can be written as

\begin{equation}
\hat{H}=\dfrac{1}{2}\sum_{\xi}\hat{p}_{\xi}^{2}+\dfrac{1}{2\hbar^{2}}\sum_{\xi\chi}\hat{q}_{\xi}\hat{q}_{\chi}K_{\xi\chi}.
\end{equation}
Here we have introduced mass-weighted displacement operators $\hat{q}_{\xi}=\sqrt{M_{\xi}}\hat{Q}_{\xi}$
and mass-scaled momentum operators $\hat{p}_{\xi}=\hat{P}_{\xi}/\sqrt{M_{\xi}}$
as conjugate variables, obeying $[\hat{q}_{\xi},\hat{p}_{\chi}]=\mbox{i}\hbar\delta_{\xi\chi}$
and $[\hat{q}_{\xi},\hat{q}_{\chi}]=[\hat{p}_{\xi},\hat{p}_{\chi}]=0$,
and $\xi=(j,c)$ denotes a Cartesian component $c=x,y,z$ of atom
$j$ at position $\vec{R}_{j}=\vec{R}_{j}^{(0)}+\vec{Q}_{j}$. The
phononic system is characterized by its dynamical matrix $K_{\xi\chi}=\hbar^{2}\mathcal{H}_{\xi\chi}/(\sqrt{M_{\xi}M_{\chi}})$,
which is the second-order mass-weighted derivative (or Hessian) of
the total ground state energy with respect to the Cartesian atomic
coordinates, $\mathcal{H}_{\xi\chi}=\partial_{\xi\chi}^{2}E$. These
harmonic force constants are computed within DFPT. 

The local displacement basis allows us to partition the dynamical
matrix into left (L), central (C), and right (R) parts
\begin{equation}
\boldsymbol{K}=\left(\begin{array}{ccc}
\boldsymbol{K}_{\mathrm{LL}} & \boldsymbol{K}_{\mathrm{LC}} & \mathrm{\boldsymbol{0}}\\
\boldsymbol{K}_{\mathrm{CL}} & \boldsymbol{K}_{\mathrm{CC}} & \boldsymbol{K}_{\mathrm{CR}}\\
\boldsymbol{0} & \boldsymbol{K}_{\mathrm{RC}} & \boldsymbol{K}_{\mathrm{RR}}
\end{array}\right).\label{eq:Kmatrix}
\end{equation}
Under the assumption of vanishing coupling between L and R, $\boldsymbol{K}_{\mathrm{RL}}=\boldsymbol{K}_{\mathrm{LR}}=\boldsymbol{0}$,
the transmission function for ballistic phonons can be written as\cite{Mingo2003,Wang2008,Asai2008}

\begin{equation}
\tau_{\textrm{ph}}(E)=\mathrm{Tr}\left[\boldsymbol{D}_{\mathrm{CC}}^{\textrm{r}}(E)\boldsymbol{\Lambda}_{\textrm{L}}(E)\boldsymbol{D}_{\mathrm{CC}}^{\textrm{a}}(E)\boldsymbol{\Lambda}_{\textrm{R}}(E)\right],
\end{equation}
where we have expressed the phonon transmission function in terms
of Green's functions solely defined on C. 

We use the general definition of the nonequilibrium phonon Green's
function on the Keldysh contour 
\begin{equation}
D_{\xi\chi}(\tau,\tau^{\prime})=-i\bigl\langle T_{C}\hat{S}_{C}\left(\hat{q}_{\xi}(\tau)\hat{q}_{\chi}(\tau^{\prime})\right)\bigr\rangle,\label{eq:D_TcSc}
\end{equation}
where the displacement operators $\hat{q}_{\xi}(\tau)$ are in the
interaction picture, $\hat{S}_{C}=\textrm{exp}[-\textrm{i}/\hbar\int_{C}\mbox{d}\tau\hat{H}_{i}(\tau)]$
is the time evolution operator on the Keldysh contour, and $T_{C}$
is the corresponding time-ordering operator.\cite{Wang2006,Wang2008,Stefanucci2013}
To calculate the phononic heat current, we need the retarded and advanced
Green's functions in energy space. They are obtained by a Fourier
transform 
\begin{equation}
D_{\xi\chi}^{\mathrm{r,a}}(E)=\dfrac{1}{\hbar}\int_{-\infty}^{\infty}\mbox{d}tD_{\xi\chi}^{\mathrm{r,a}}(t)\mbox{e}^{iEt/\hbar}
\end{equation}
of the real-time functions, derived from Eq.~(\ref{eq:D_TcSc}) as\cite{Wang2006,Wang2008}

\begin{flalign}
\boldsymbol{D}_{\mathrm{CC}}^{\mathrm{r}}(E) & =\Bigl[\left(E+\textrm{i}\eta\right)^{2}\boldsymbol{1}-\boldsymbol{K}_{\textrm{CC}}-\boldsymbol{\Pi}_{\textrm{L}}^{\mathrm{r}}(E)\Bigr.\nonumber \\
 & \qquad\Bigl.-\boldsymbol{\Pi}_{\textrm{R}}^{\mathrm{r}}(E)\Bigr]\label{eq:Dr(E)}
\end{flalign}
with $\boldsymbol{D}_{\mathrm{CC}}^{\mathrm{a}}(E)=\boldsymbol{D}_{\mathrm{CC}}^{\mathrm{r}}(E)^{\dagger}$
and an infinitesimal quantity $\eta>0$. For non-interacting phonons
in the electrodes the corresponding contact self-energies can be calculated
exactly as

\begin{equation}
\boldsymbol{\Pi}_{Y}^{\mathrm{r}}(E)=\boldsymbol{K}_{\textrm{C}Y}\boldsymbol{d}_{YY}^{\mathrm{r}}(E)\boldsymbol{K}_{Y\mathrm{C}}.
\end{equation}
Here, $\boldsymbol{d}_{YY}^{\mathrm{r}}(E)$ is the surface Green's
function of lead $Y=\{\textrm{L, R}\}$. Additionally, we have defined
the spectral density of the electrodes as
\begin{equation}
\boldsymbol{\Lambda}_{Y}=\textrm{i}(\boldsymbol{\Pi}_{Y}^{\mathrm{r}}-\boldsymbol{\Pi}_{Y}^{\mathrm{a}})
\end{equation}
with $\boldsymbol{\Pi}_{Y}^{\textrm{a}}(E)=\boldsymbol{\Pi}_{Y}^{\textrm{r}}(E)^{\dagger}$.

\begin{figure}
\centering{}\includegraphics[width=1\linewidth]{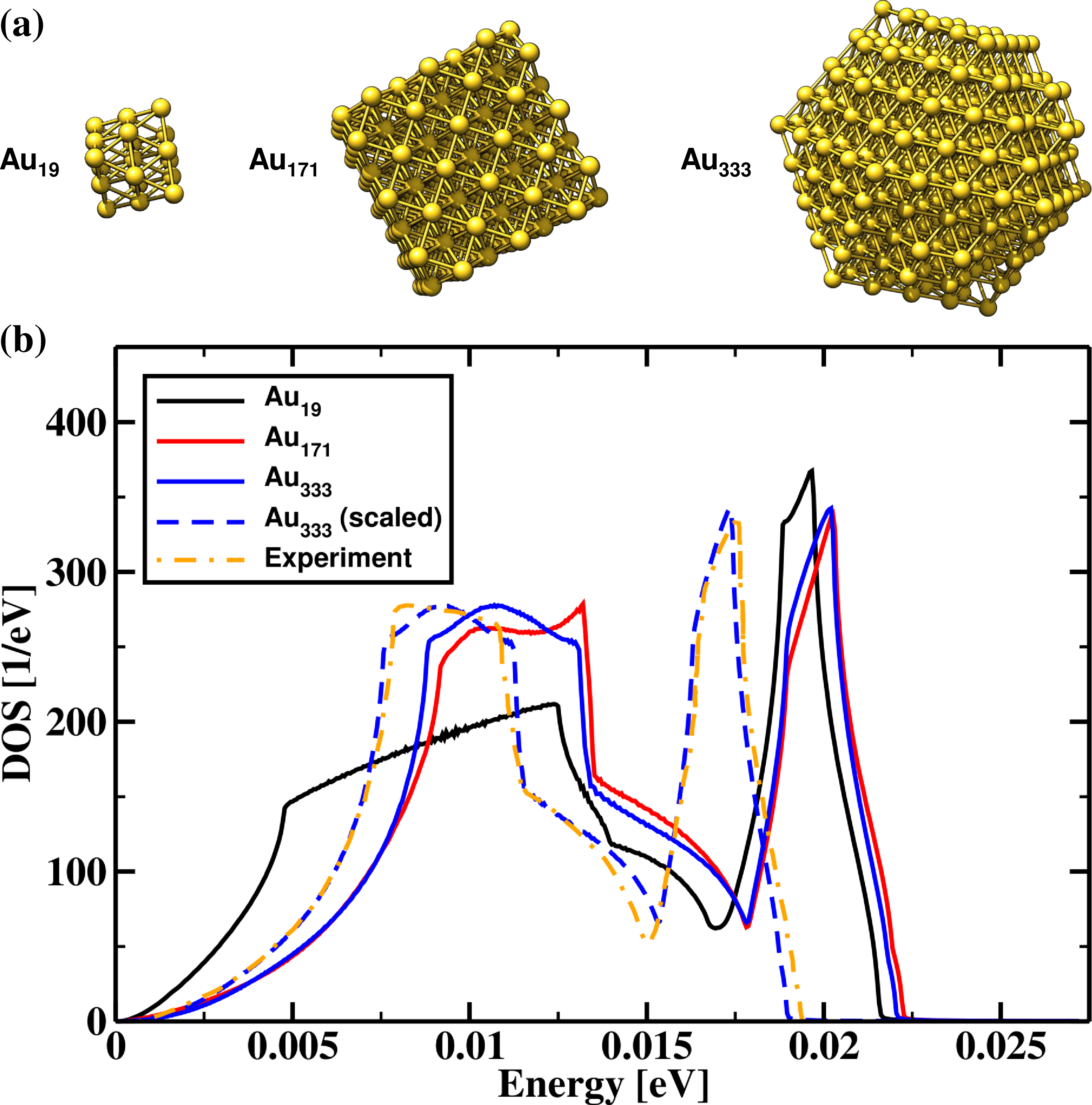}\protect\caption{(a) Shape of the three clusters, $\mathsf{Au}_{19}$, $\mathsf{Au}_{171}$
and $\mathsf{Au}_{333}$, used to extract the bulk parameters. (b)
Bulk phonon density of states calculated with parameters extracted
from the three cluster shown in (a) and experimental DOS from Ref.~\citenum{Lynn1973}.
The blue dashed curve is the DOS as obtained from the $\mathsf{Au}_{333}$
cluster scaled by 0.85 to fit the experimental vibration energies.\label{fig:clusterdos} }
\end{figure}

Similar to our charge quantum transport approach\cite{Pauly2008}
we employ also for the phononic case a cluster-based procedure. We
calculate the dynamical matrix $\boldsymbol{K}_{\mathrm{ECC}}$ of
a large, but finite cluster consisting of the molecule of interest
and parts of the electrodes. Using the partitioning of the ECC, given
in Eq.~(\ref{eq:Kmatrix}) and depicted in Fig.~\ref{fig:mols_geos}b,
we extract $\boldsymbol{K}_{\mathrm{CC}}$, $\boldsymbol{K}_{\mathrm{LC}}$
and $\boldsymbol{K}_{\mathrm{RC}}$ from $\boldsymbol{K}_{\mathrm{ECC}}$.
What remains to be calculated to determine $\tau_{\mathrm{ph}}(E)$
are the surface Green's functions of the electrodes $\boldsymbol{d}_{YY}^{\mathrm{r}}(E)$.
They are constructed for perfect semi-infinite crystals. The dynamical
matrix of the \textsf{Au} electrode is derived from those of a spherical
cluster, see Fig.~\ref{fig:clusterdos}a. As in the crystal the atoms
of the \textsf{Au} cluster are positioned on an fcc lattice with lattice
constant $a_{0}=4.08$~Å. We exploit that for large enough clusters
the force constants from the central atom to its neighbors will be
similar to the bulk. Following the procedure introduced in Ref.~\citenum{Pauly2008}
for the charge transport, we construct the dynamical matrix of the
crystalline electrode via these extracted bulk couplings. From this
we compute surface and bulk Green's functions, $\boldsymbol{d}_{YY}^{r}(E)$
and $\boldsymbol{d}_{\textrm{Bulk}}^{r}(E)$, respectively, iteratively
by means of the decimation technique.\cite{Guinea1983,Pauly2008,Wang2008} 

To check the quality of the extracted bulk parameters, we compare
the bulk phonon density of states (DOS) against available experimental
data. It can be directly obtained from the retarded bulk Green's function
\begin{equation}
\rho(E)=-\dfrac{2}{\pi}E\mbox{Im}\left(\mbox{Tr}\left[d_{\textrm{Bulk}}^{\mathrm{r}}(E)\right]\right).
\end{equation}
The comparison of the DOS, calculated from parameters extracted from
$\mathsf{Au}_{19}$, $\mathsf{Au}_{171}$ and $\mathsf{Au}_{333}$
clusters, with the experimental data of Ref.~\citenum{Lynn1973}
is shown in Fig.~\ref{fig:clusterdos}b. For the smallest cluster
$\mathsf{Au}_{19}$ the calculated DOS already resembles the general
features of the experimental one. However, in the lower energy range
from 0 to 15 meV the agreement is still poor. Increasing the cluster
size to 171 atoms leads to a change especially in the low-energy range
of the DOS. If we increase the cluster size further to 333 atoms,
we observe only small modifications. Comparing the DOS calculated
for the parameters extracted from the $\mathsf{Au}_{333}$ cluster
with the experiment,\cite{Lynn1973} we see that the DFT results resemble
the experimental DOS closely, but overestimate the vibrational energies
by around 15\%. By scaling the energy axis of the calculated DOS by
a factor of 0.85, we basically obtain a perfect match with the experimental
results. Regarding the scaling, we note that in principle both the
$x$- and $y$-axis of the DOS plot need to be scaled simultaneously
with inverse factors to conserve the energy integral of the DOS. But
since the experimental data is measured in arbitrary units, a scaling
factor for the $y$ axis is already used. 

We want to point out that, in order to obtain a reliable DOS as well
as phonon transport properties, it is crucial to choose the broadening
factor $\eta$ sufficiently small. Otherwise the sharp features in
the spectra are smeared out. Here, we used a broadening of $\eta=6.8\,\mu\textrm{eV}$,
for which we obtained both a converged DOS and converged transmission
spectra with respect to $\eta$. Due to the small $\eta$ a large
number of transverse $k$-points has to be chosen in the decimation
procedure. We found the results to be converged for a $k$-point sampling
of $512\times512$. After the DFPT step, this large number of $k$-points
is the bottle neck of the transport calculations. However, the calculation
can be parallelized with respect to the $k$-points and after transforming
the surface and bulk Green's functions back to real space, they can
be conveniently stored on a hard drive and reused in later calculations.

\textcolor{red}{}

\subsection{Thermoelectric figure of merit}

The common measure for thermoelectric efficiency is given by the figure
of merit 
\begin{equation}
ZT=\dfrac{GS^{2}}{\kappa_{\textnormal{el}}+\kappa_{\textnormal{ph}}}T,\label{eq:ZT_def}
\end{equation}
which is determined from the thermoelectric transport coefficients
in Eqs.~(\ref{eq:G})-(\ref{eq:kel}) and (\ref{eq:kph}) in the
linear response regime. An upper bound for $ZT$ in the limit of vanishing
phonon thermal transport $\kappa_{\textnormal{ph}}\rightarrow0$ is
given by the purely electronic contribution 
\begin{equation}
Z_{\textnormal{el}}T=\frac{S^{2}G}{\kappa_{\textnormal{el}}}T=\frac{S^{2}}{L}.\label{eq:ZTel_def}
\end{equation}
Here, we have introduced the Lorenz number $L=\kappa_{\textnormal{el}}/GT$.
With $Z_{\textrm{el}}T$ we can express the figure of merit in a slightly
different form as 
\begin{equation}
ZT=\dfrac{Z_{\mathrm{el}}T}{1+\kappa_{\textnormal{ph}}/\kappa_{\textnormal{el}}}.\label{eq:ZTwithZeT}
\end{equation}

\section{Results and discussion}

\begin{figure}[!tb]
\centering{}\includegraphics[width=1\linewidth]{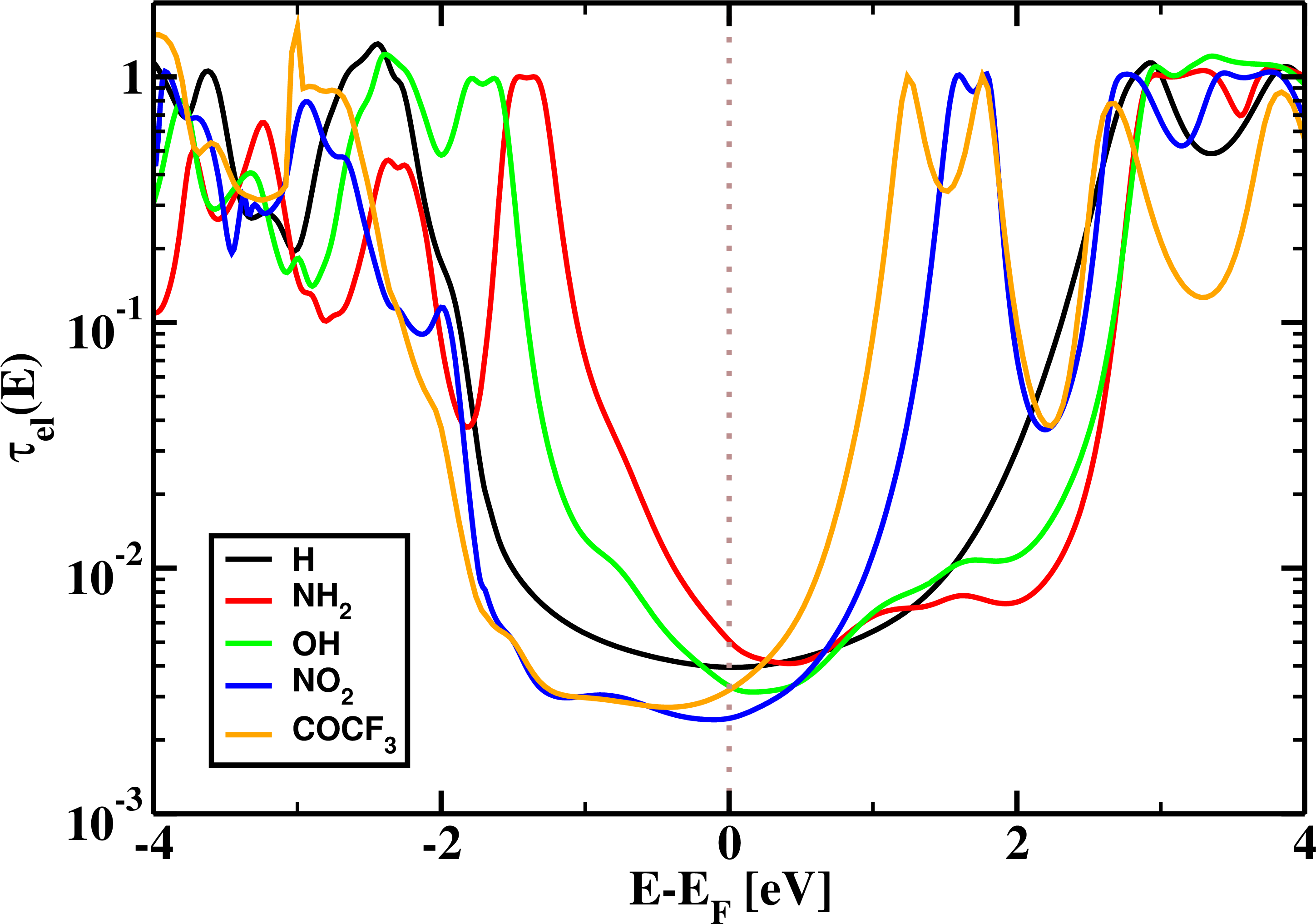}\protect\protect\caption{Energy-dependent electronic transmission probability $\tau_{\mathrm{el}}(E)$
relative to $E_{F}$.\label{fig:el_trans}}
\end{figure}

For the unsubstituted paracyclophane molecule ($\mathsf{X=H}$), $E_{\mathrm{F}}$
is located almost in the middle of the HOMO-LUMO gap (Fig.~\ref{fig:el_trans}).
The conductance is found to be $G_{\textrm{DFT+\ensuremath{\Sigma}}}=0.004\,G_{0}$
(without self-energy correction $G_{\textrm{DFT}}=0.07\,G_{0}$),
which compares reasonably well with the experimental data $G_{\textrm{exp}}\approx0.01\,G_{0}$,
given in Ref.~\citenum{Schneebeli2011}. Introducing the substituents
leads to a shift of the molecular orbital energies relative to $E_{\mathrm{F}}$.\cite{Buerkle2012}
For the electron-donating substituents $\mathsf{NH}_{2}$ and $\mathsf{OH}$,
the molecular frontier orbital energies increase due to the increased
Coulomb repulsion in the conjugated $\pi$-electron system of the
benzene rings. This moves the HOMO resonance around 1.1 eV closer
to $E_{\mathrm{F}}$. The effect is slightly larger for $\mathsf{NH}_{2}$
than for $\mathsf{OH}$ (Fig.~\ref{fig:el_trans}). The LUMO level
is largely unaffected, which suggests that the two substituents mainly
couple to the occupied $\pi$-orbitals. For the electron-withdrawing
substituents $\mathsf{NO_{2}}$ and $\mathsf{COCF_{3}}$, we observe
the opposite effect. The decreased Coulomb repulsion moves the LUMO
resonance around 1.5 eV towards $E_{\mathrm{F}}$, with a slightly
larger shift for $\mathsf{COCF_{3}}$ (Fig.~\ref{fig:el_trans}).

\begin{figure}[!tb]
\centering{}\includegraphics[width=1\linewidth]{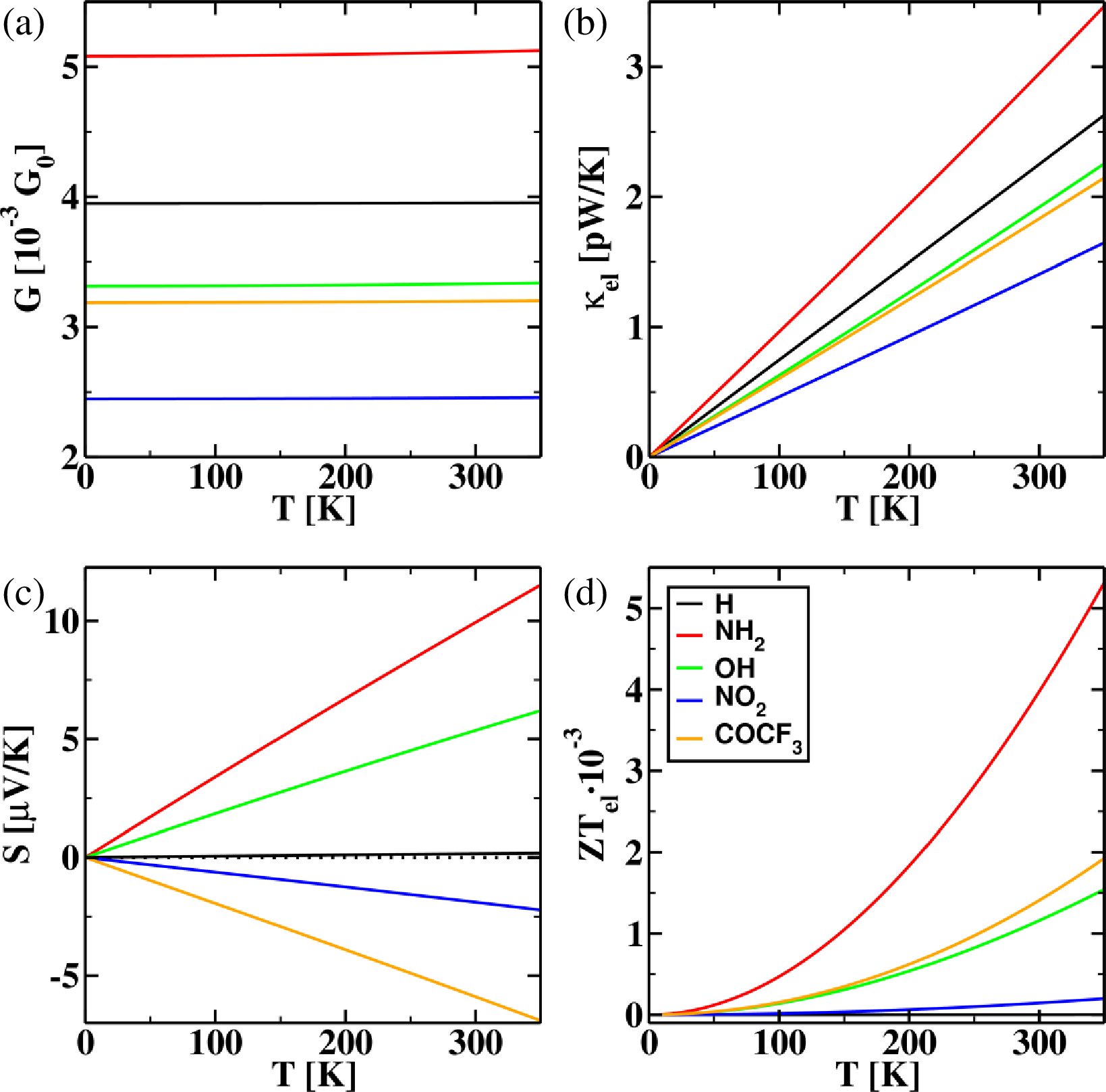}\protect\protect\caption{Temperature dependence of the thermoelectric transport coefficients,
namely (a) electric conductance, (b) electron thermal conductance,
and (c) thermopower. (d) Electronic contribution $Z_{\mathrm{el}}T$
to the figure of merit.\label{fig:thermopower_zet}}
\end{figure}
\begin{figure*}[!]
\begin{centering}
\includegraphics[width=1\linewidth]{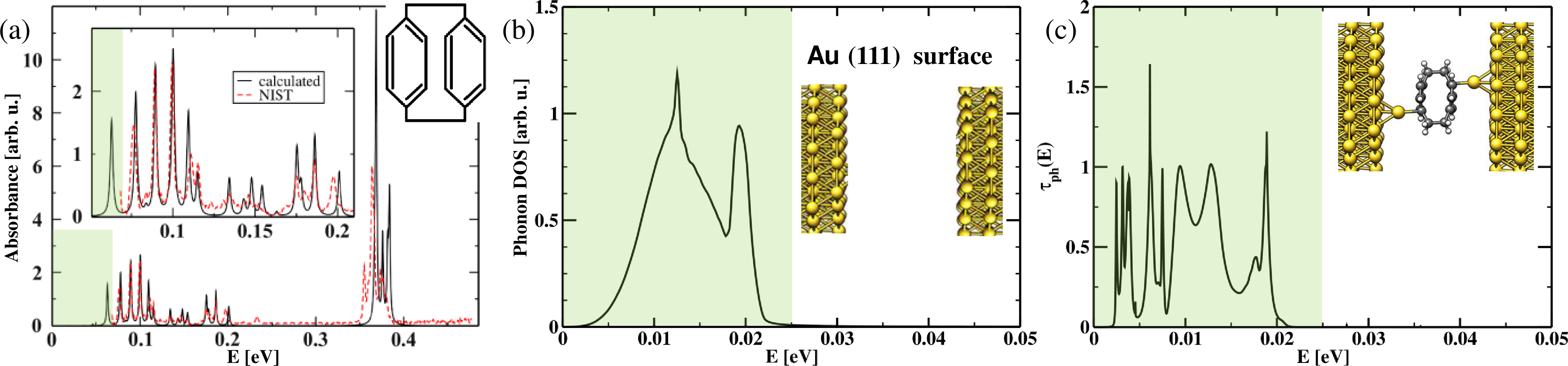} 
\par\end{centering}

\centering{}\protect\protect\caption{(a)\textbf{ }Calculated and measured\cite{nist2014} infrared spectra
of {[}2,2{]}paracyclophane in the gas phase, showing the infrared-active
molecular vibrations. (b) Phonon DOS of a \textsf{Au }(111) surface.
(c) Phonon transmission $\tau_{\mathrm{ph}}(E)$ for {[}2,2{]}paracyclophane
connected to \textsf{Au} electrodes. Green-shaded areas indicate the
energy window, where the electrode phonon DOS is finite.\label{fig:molelectrovibstrans}}
\end{figure*}

In Fig.~\ref{fig:thermopower_zet}a-c the temperature dependence
of the thermoelectric transport coefficients is summarized. Before
discussing the individual features, it is worth stressing that due
to the off-resonant transport and smooth transmission functions around
$E_{\mathrm{F}}$, the temperature dependence of the thermoelectric
transport coefficients (Eqs.~(\ref{eq:G})-(\ref{eq:kel})) is dominated
by the lowest order of the Sommerfeld expansion of Eq.~(\ref{eq:Kn}).\cite{Ashcroft1976,Gomez-Silva2012}

Despite the substituent-induced movement of the orbital energies,
the transmission at $E_{\textrm{F}}$ remains roughly constant as
compared to the unsubstituted molecule. Hence the influence of the
substituents on the electric conductance $G$ as well as the electron
thermal conductance $\kappa_{\textrm{el}}\approx LTG$ remains small.
For $\mathsf{NH}_{2}$, $G$ and $\kappa_{\textrm{el}}$ are slightly
increased. For the other three substituents a decrease is observed
(Fig.~\ref{fig:thermopower_zet}a,b). However, the slope of the transmission
function at $E_{\textrm{F}}$, and hence the thermopower $S$, changes
substantially. For the unsubstituted molecule the transmission function
is flat around $E_{\textrm{F}}$, resulting in an almost vanishing
thermopower (Fig.~\ref{fig:thermopower_zet}c). On the other hand,
for all four substituents the absolute value of $S$ is largely increased
(Fig.~\ref{fig:thermopower_zet}c). Moreover the electron-donating
substituents $\mathsf{NH}_{2}$ and $\mathsf{OH}$ tune the slope
of $\tau_{\mathrm{el}}(E)$ at $E_{\mathrm{F}}$ to be negative, giving
a positive thermopower, characteristic for HOMO-dominated transport
($p$-type).\cite{Paulsson2003,Baheti2008,Tan2011,Balachandran2013}
In contrast, the electron-withdrawing substituents give rise to LUMO-dominated
transport ($n$-type), and hence the thermopower is negative (Fig.~\ref{fig:thermopower_zet}c).

\begin{figure}[!tb]
\begin{centering}
\includegraphics[width=1\linewidth]{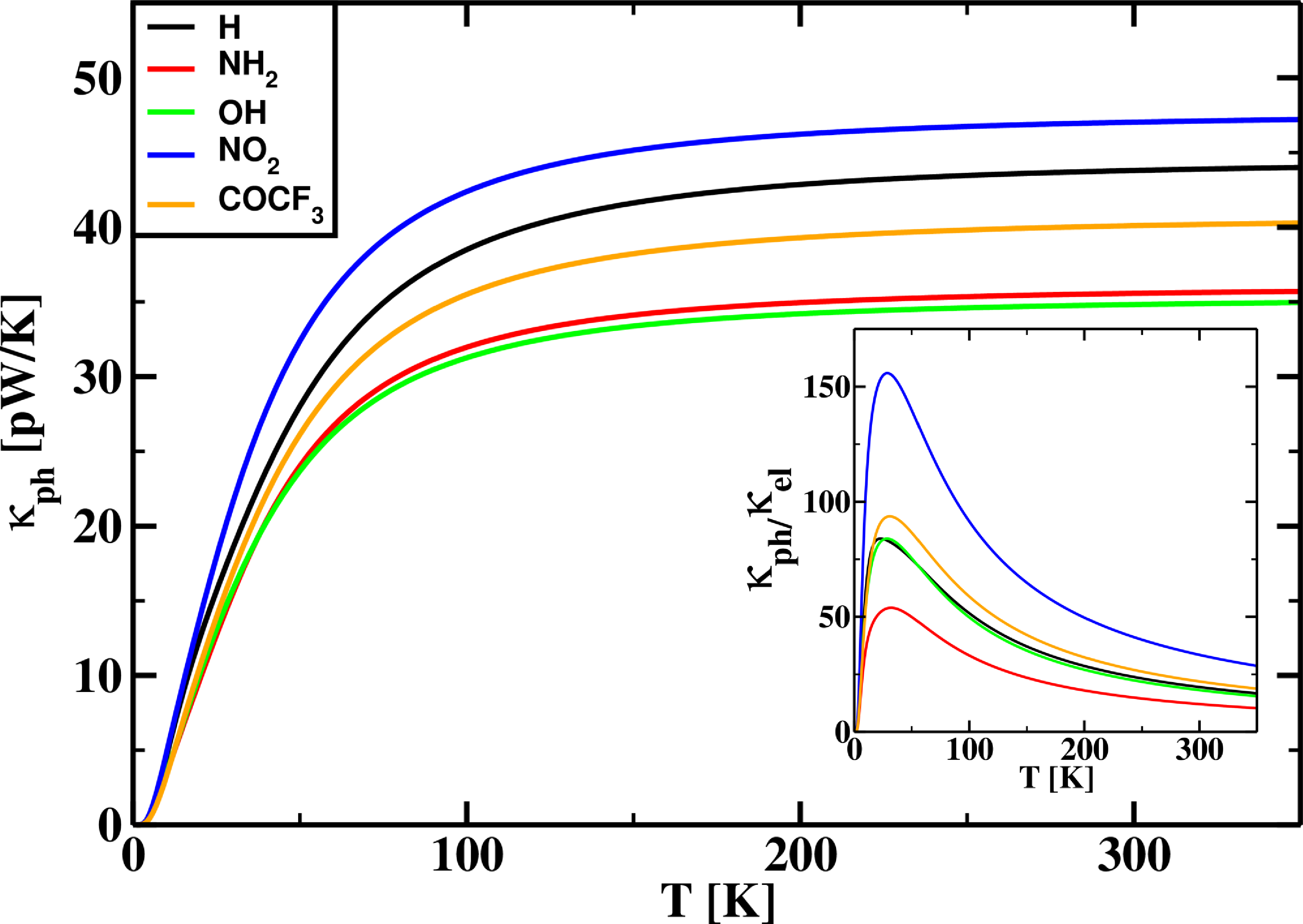} 
\par\end{centering}

\centering{}\protect\protect\caption{Temperature-dependent phonon thermal conductance. Inset: Ratio $\kappa_{\mathrm{ph}}/\kappa_{\mathrm{el}}$
between phonon and electron thermal conductance.\label{fig:kappa_ph-1} }
\end{figure}

Ultimately, we can combine the thermoelectric transport coefficients
to the electronic contribution to the figure of merit $Z_{\textrm{el}}T$,
which is displayed in Fig.~\ref{fig:thermopower_zet}d. Due to the
almost vanishing thermopower of the unsubstituted molecule, $Z_{\textrm{el}}T$
remains very small. On the other hand, due to the quadratic dependence
of $Z_{\textrm{el}}T$ on $S$, the enhanced thermopower for the substituted
molecules increases $Z_{\mathrm{el}}T$ largely. The $\mathsf{NH}_{2}$-substituted
molecule shows the highest $Z_{\textrm{el}}T$ of around $4\cdot10^{-3}$
at room temperature ($T=300\,\textrm{K}$). Overall $Z_{\textrm{el}}T$
remains rather small for the {[}2,2{]}paracyclophane derivatives studied
here. The fact that it is possible to tune the system to be either
$p$- or $n$-type by simply replacing the substituents is very encouraging,
as this simple approach can be easily transferred to other $\pi$-conducting
systems.

Now we turn to the phononic contribution to the thermal conductance,
which is often neglected in the calculations or treated in an approximate
manner. Here, however, we determine also the phononic thermal conductance
accurately from first-principles, as described in Sec.~\ref{sub:Phonon-transport}.
Compared to the molecular vibration spectrum (Fig.~\ref{fig:molelectrovibstrans}a),
the electrode phonon density of states (DOS) is finite only in a narrow
energy window (Fig.~\ref{fig:molelectrovibstrans}b). Even if the
presence of the electrodes can rather strongly renormalize the molecular
vibration spectrum and the character of the individual modes, simply
speaking, just those molecular vibrations within the energy window
defined by the electrode phonon DOS can contribute to the phonon transport
(Fig.~\ref{fig:molelectrovibstrans}c).

The temperature dependence of the phonon thermal conductance $\kappa_{\mathrm{ph}}$
is displayed in Fig.~\ref{fig:kappa_ph-1}. We obtain the expected
ballistic behavior with a steep increase for low temperatures, as
the phonon modes are getting occupied, while for higher temperatures
$\kappa_{\mathrm{ph}}$ approaches a constant value. Substituting
the paracyclophane has a twofold effect. Firstly it influences the
molecular vibration spectrum directly, moving modes in and out of
the energy window defined by the electrode phonon DOS. Secondly, it
changes the binding geometry between the molecule and the \textsf{Au}
electrode, modifying at the same time the coupling between the molecular
vibrations and the electrode phonons. Both effects are, however, difficult
to separate. Substituting with $\mathsf{NO_{2}}$ increases the phonon
thermal conductance, while the other three substituents decrease it
compared to the unsubstituted molecule. Eventually $\kappa_{\mathrm{ph}}$
remains comparable for all five molecules (Fig.~\ref{fig:kappa_ph-1})
and is, in the temperature range $10\,\textrm{K}<T<400\,\textrm{K}$,
for all molecules at least one order of magnitude larger than the
electronic thermal conductance (inset of Fig.~\ref{fig:kappa_ph-1}).

\begin{figure}[!tb]
\centering{}\includegraphics[width=0.8\linewidth]{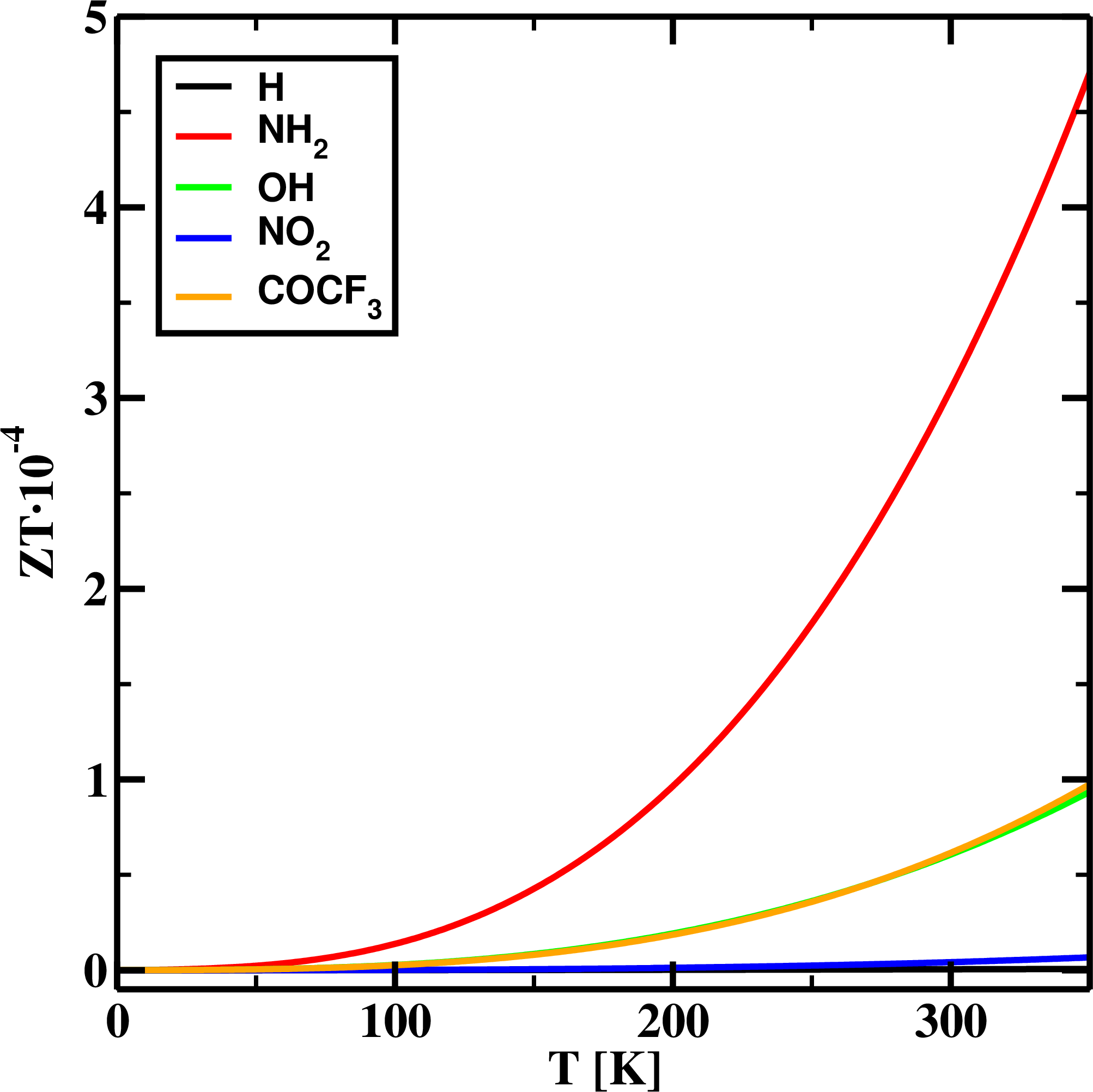}\protect\protect\caption{Temperature dependence of the figure of merit $ZT$, including both
electronic and phononic linear response transport coefficients.\label{fig:ZT}}
\end{figure}

Since $\kappa_{\mathrm{ph}}/\kappa_{\mathrm{el}}>10$ for $10\,\textrm{K}<T<400\,\textrm{K}$,
$ZT$ is strongly suppressed according to Eq.~(\ref{eq:ZTwithZeT})
(Fig.~\ref{fig:ZT}). The largest $ZT$ is obtained for $\mathsf{NH_{2}}$
substitution with $ZT=4\cdot10^{-4}$ at room temperature ($T=300\,\mathrm{K}$).
The $\mathsf{OH}$ and $\mathsf{COCF_{3}}$ substituted paracyclophanes
have almost the same $ZT$ in the whole temperature range with $ZT=0.9\cdot10^{-4}$
at $T=300\,\mathrm{K}$, but the former one is hole-conducting while
the latter one is electron-conducting.

\section{Conclusions}

For a series of paracyclophane-based single-molecule junctions we
computed the electrical and heat transport properties fully from first-principles.
Combining $\mathrm{DFT}+\Sigma$-based electronic structure and DFPT-derived
harmonic force constants with the Green's function formalism and the
Landauer approach allowed us to make quantitative predictions of all
linear response transport coefficients relevant for the thermoelectric
figure of merit $ZT$.

We showed that the transport properties can be tuned by functionalizing
the paracyclophane molecule appropriately. The main influence of the
functional groups was found to be on the junction thermopower, where
both its absolute value and sign are affected. This is due to the
fact that the Fermi energy lies right in the middle of the molecular
HOMO-LUMO gap of the bare paracyclophane molecule, leading to a vanishing
thermopower and consequently $ZT$. The electron-donating substituents
$\mathsf{NH}_{2}$ and $\mathsf{OH}$ tune the thermopower to be positive,
while the electron-withdrawing groups $\mathsf{NO_{2}}$ and $\mathsf{COCF_{3}}$
give rise to a negative value. The largest enhancement of the absolute
value of the thermopower was found for $\mathsf{NH}_{2}$, followed
by $\mathsf{OH}$ and $\mathsf{COCF_{3}}$. The latter two substituents
yield a comparable absolute value of $S$, but a different sign.

Naturally, the non-vanishing thermopower increases the electronic
contribution $Z_{\mathrm{el}}T$ to the thermoelectric figure of merit
largely. However, for the studied off-resonant conduction, the phonon
thermal conductance is, even at room temperature, at least one order
of magnitude larger than the electronic one, leading to a significant
suppression of the overall $ZT$. Therefore we need to stress that
it is necessary to include the phonon contribution to the thermal
conductance for low-conducting molecular junctions in order to obtain
accurate and reliable predictions of $ZT$.

In our present study the largest $ZT$ was found for the $\mathsf{NH_{2}}$-functionalized
paracyclophane with $ZT=4\cdot10^{-4}$ at $T=300\,\mathrm{K}$. Even
if the $ZT$ remains rather small, we could demonstrate for a series
of realistic molecules that it is possible to chemically adjust the
transport properties and to enhance the thermoelectric figure of merit
as well as to realize $p$- and $n$-conducting junctions using the
same molecular framework.

In the future more complex molecular structures and different functional
groups could be used to enhance the thermoelectric figure of merit
further. Beside the chemical tuning of mainly the electronic transport
properties, which was demonstrated in this work, limiting the phonon
contribution to the thermal current is necessary to improve $ZT$.
Nonlinear effects, leading to increased phonon-phonon scattering,
could provide a convenient way to suppress the phonon transport. Such
a discussion will be the topic of future work. 
\begin{acknowledgments}
MB and YA acknowledge fruitful discussions with H.~Nakamura. This
work was partly supported by a FY2012 (P12501) Postdoctoral Fellowship
for Foreign Researchers from the Japan Society for Promotion of Science
(JSPS) and by a JSPS KAKENHI, i.e.~‘Grant-in-Aid for JSPS Fellows’,
grant no.~24·02501. YA is also thankful to another KAKENHI, i.e.~Grant-in-Aid
for Scientific Research on Innovation Areas ``Molecular Architectonics:
Orchestration of Single Molecules for Novel Functions'' (\#25110009).

TJH would like to thank the Karlsruhe House of Young Scientists for
financial support during his stay at the Nanosystem Research Institute.

FP gratefully acknowledges funding from the Carl Zeiss foundation
and the Junior Professorship Program of the Ministry for Science,
Research, and Art of the state of Baden-Württemberg. 
\end{acknowledgments}

\bibliographystyle{apsrev4-1}
\bibliography{paracyclophane_prb}

\end{document}